\documentclass[epj]{svjour}
\usepackage{epsfig}

\begin{document}
\title{
Competition of disorder and interchain hopping 
 in a two-chain Hubbard model
}
\author{E. Orignac\inst{1} \and Y. Suzumura\inst{2,3} }
\institute{LPTENS CNRS-UMR 8549 24, Rue
Lhomond 75231 Paris Cedex 05, France \thanks{\email{orignac@lpt.ens.fr}}
\and Department of Physics, Nagoya University, Nagoya 464-8602 ,
Japan  \and
CREST, Japan Science and Technology Corporation (JST), Japan }
\date{\today}
\titlerunning{Disorder in a Hubbard ladder}
\authorrunning{Orignac and Suzumura}
\abstract{
 We study the interplay of Anderson localization and interaction in a two 
 chain Hubbard ladder allowing for arbitrary ratio of disorder strength
 to interchain coupling. We obtain three different types of 
 spin gapped localized phases depending on the strength of disorder: 
a pinned  $4k_F$ Charge Density Wave (CDW) for weak disorder, a  pinned $2k_F$
 CDW$^\pi$ for intermediate disorder and two independently pinned single
 chain $2k_F$ CDW for strong disorder. 
 Confinement of electrons can be obtained  as a result of strong disorder 
 or strong attraction. We give the full phase diagram as a function of
disorder, interaction strength and interchain hopping. We also study
the influence of interchain hopping on localization length and show
that localization is enhanced by a small interchain hopping but
suppressed by a large interchain hopping. 
}
\maketitle

\section{Introduction}\label{sec:intro}

In the recent years, the two-chain Hubbard model received a lot of
theoretical attention both analytically
\cite{rice_srcuo,fabrizio_2ch_rg,nagaosa_2ch,finkelstein_2ch,kveschenko_spingap,schulz_2chains,nagaosa_chiral_anomaly_1d,balents_2ch,shelton_tj_ladder,yoshioka_2ch,yoshioka_coupledchains_interaction} 
and numerically
\cite{poilblanc_2ch_mc,hayward_2chain,tsunegutsu_2ch,poilblanc_2ch,fabrizio_q1d,noack_dmrg_2ch,dagotto_lanczos_2ch,park_2ch_dmrg}.
The outcome of these studies was that the two-chain Hubbard model with
repulsive interactions would present a metallic phase with a spin gap
containing dominant ``d-wave'' superconducting fluctuations, a
situation very reminiscent of the situation in two-dimensional cuprate
superconductors. This result leads to the conjecture that the two-chain
Hubbard model could be considered as a toy-model of cuprate
superconductivity. More recently, experiments revealed that the material 
with two-chain structure $\mathrm{Sr_{14}Cu_{24}O_{41}}$ doped with
$\mathrm{Ca}$ could become metallic\cite{uchara_SrCaCuO} and displayed a 
superconducting phase under pressure. This superconducting phase has been interpreted
as the stabilization of ``d-wave'' fluctuations by interladder Josephson 
coupling. 
Such an interpretation is not clear enough because  doping by
$\mathrm{Ca}$ is not only introducing carriers in the system, but also
inducing disorder. 

In a one-dimensional system, even an
infinitesimal disorder induces Anderson localization and  an
insulating state   in the presence of
interactions\cite{giamarchi_loc}. 
Therefore  it is crucial to 
 understand the effect of interchain
  coupling on Anderson localization.  At present,  the
effect of an infinitesimal disorder in the limit of strong interchain
coupling has been
analyzed\cite{orignac_2chain_short,orignac_2chain_long}. 
The outcome of
this study is that although the presence of the spin-gap suppresses the
coupling of disorder to $2k_F$ fluctuations, disorder can still couple
to $4k_F$ fluctuations ($k_{F}$ being the Fermi wave vector)
 leading to an insulating phase rather than a 
superconducting phase at low temperature. Indeed, there is some
experimental evidence from transport measurements \cite{akimitsu_mit_srcacocuo,osafune_srcacuo_cdw,adachi_ladder_loc,ruzicka_ladder_electrodynanics}
that the phase at low temperature under ambient
pressure is a pinned Charge Density Wave (CDW) state. 
 This, as well as the disappearance of spin gap in NMR and the
reduction of anisotropy of electrical conductivity close to the
superconducting transition has lead to the idea 
that pressure could  cause a dimensional crossover from 1D insulator 
 to 2D superconductor 
 and that superconductivity in this system is
  intrinsically two-dimensional
phenomenon\cite{mayaffre_supra_echelles,nagata_ladder_single,nagata_ladder_pressure,piskunov_ladder_nmr}.
Building on the results of \cite{orignac_2chain_long}, a
phenomenological theory of transport in the insulating state of the
ladder system was proposed in \cite{arrigoni_2ch_impurity}.   

Besides its relevance to  ladder compounds, the study of two-chain
systems is also important for the study of transport 
in systems such as metallic carbon
nanotubes\cite{ijima_nanotubes_synthesis,odintsov_nanotubes} and quantum
wires\cite{meirav_gaas_1d,meirav_wires,tarucha_wire_1d,starykh_quantum_wires}.
 It has also a
theoretical interest in itself since the interplay of localization and
interaction in dimension higher than one remains poorly
understood\cite{belitz_localization_review} and coupled quasi-one
dimensional system may allow for a controlled study of the effect of
increasing space dimensionality on disorder and interaction. 

In the present paper, we extend the work of
Refs. \cite{orignac_2chain_short,orignac_2chain_long} to arbitrary
ratio of disorder to interchain coupling. A study in the case of
spinless fermions revealed an interesting transition from two chain to
single chain behavior as disorder strength was increased
\cite{orignac_spinless_ladder}, indicating 
  confinement of fermions by disorder.
 We shall analyze the two chain system  with disorder 
 by bosonization\cite{solyom_revue_1d,emery_revue_1d,fukuyama_takayama,schulz_houches_revue,voit_bosonization_revue} and 
RG (renormalization group) 
 techniques and discuss the different disordered phases that are 
 obtained as well as  the confinement phase
\cite{suzumura_tsuchiizu_gruner,tsuchiizu_confinement_ladder_long}. 
 We use the same techniques as Kawakami and Fujimoto
\cite{fujimoto_mott+disorder_2ch}, who  
analyzed the interplay of disorder and
Umklapp scattering in the two chain ladder at commensurate filling
in the regime in which interchain hopping is
dominant\cite{fujimoto_mott+disorder_2ch}.
However  
 we examine  a different 
system at incommensurate filling, but  allow disorder to become
stronger than interchain hopping. Other related work in the ladder
considered the
effect of forward scattering  on the renormalization of disorder
\cite{mori_ladders_loc,mori_persistent_2ch,sandler_disorder_qwires} 
neglecting the influence of gap openings on localization
properties. Such an approximation is restricted to the limit of high
temperature or short system, and can be recovered from the general RGE
 (renormalization group equation)
of the present article.    

The plan of the paper is as follows. In Sec.~\ref{sec:hamiltonian}, we
will recall the derivation of the bosonized Hamiltonian of the
system. Then, in Sec.~\ref{sec:renorm} we will show the RG equations of
the system  
 while those of  the limit of large interchain hopping is 
 derived in Sec.~4 to compare with other related work.  
  In Sec.~\ref{sec:results}, we will demonstrate  several types 
 of spin gapped phases based on  the
outcome of RG.
Finally discussion is given in Sec.~\ref{sec:discussion}. 

\section{Hamiltonian}\label{sec:hamiltonian}

The original lattice Hamiltonian of two-chain Hubbard model with 
 disorder  reads:
\begin{eqnarray}\label{eq:lattice_hamiltonian}
H= 
 &-&t \sum_{i,p,\sigma}(c^\dagger_{i,p,\sigma}c_{i+1,p,\sigma}
    +{\rm H. c.}) 
   -t_\perp \sum_{i,\sigma}
 ( c^\dagger_{i,1,\sigma} c_{i,2,\sigma} + {\rm H.c.})  \nonumber \\ 
     &+& U \sum_{i,p} n_{i,p,\uparrow}n_{i,p,\downarrow} 
     + \sum_{i,p,\sigma} \epsilon_{i,p} n_{i,p,\sigma} ,
\end{eqnarray}
 where $n_{i,p,\sigma}=c^\dagger_{i,p,\sigma}c_{i,p,\sigma}$ and
              $\overline{ \epsilon_{i,p} \epsilon_{i',p'}}=W
                     \delta_{i,i'}\delta_{p,p'}$. 

We turn to the bonding antibonding band representation:
\begin{eqnarray}
c_{i,0,\sigma}=\frac{c_{i,1,\sigma}+c_{i,2,\sigma}}{\sqrt{2}}, \\
c_{i,\pi,\sigma}=\frac{c_{i,1,\sigma}-c_{i,2,\sigma}}{\sqrt{2}}, 
\end{eqnarray}
 and use the continuum limit to  apply
 bosonization
 techniques\cite{schulz_2chains,orignac_2chain_long,mori_persistent_2ch}. 
 Thus   the Hamiltonian can 
be rewritten as:
\begin{eqnarray}\label{eq:bosonized_hamiltonian}
H  &= &H_{{\rm pure}}+H_{{\rm impurity}},  \\
    H_{{\rm pure}}&=&\sum_{\nu=\rho,\sigma \atop r 
       =\pm} \int \frac{dx}{2\pi} \left[u_{\nu } K_{\nu r}(\pi \Pi_{\nu r})^2 
      + \frac{u_{\nu }}{K_{\nu r}} (\partial_x \phi_{\nu r})^2\right] \nonumber \\
   &+& \frac {2t_\perp}{\pi}\int dx \partial_x \phi_{\rho -} \nonumber \\
  + \frac {v_F}{2\pi a^2} & & \int dx 
  \left[ y_1 \cos 2\theta_{\rho -} \cos 2\phi_{\sigma+} 
        + y_2 \cos 2\theta_{\rho -} \cos 2 \theta_{\sigma
-}\right. \nonumber \\  
    & & \left.     + y_3 \cos 2\theta_{\rho -}\cos 2 \phi_{\sigma -}
  + y_4\cos 2\theta_{\rho -}\cos 2\phi_{\rho -} \right. \nonumber \\  
    & & \left.   
            +y_5 \cos 2\phi_{\rho -}\cos 2\phi_{\sigma+} 
            +y_6 \cos 2\phi_{\rho -} \cos 2 \theta_{\sigma -}\right.  \nonumber \\ 
  & & \left.   +y_7   \cos 2\phi_{\rho -} \cos 2 \phi_{\sigma -}
              +y_8\cos 2 \theta_{\sigma -} \cos 2 \phi_{\sigma -} 
           \right.  \nonumber \\ 
  & & \left.
              +y_9\cos 2\phi_{\sigma+} \cos 2 \phi_{\sigma -} + y_{10}\cos 2\phi_{\sigma+} \cos 2 \theta_{\sigma -}  \right], \\
  H_{{\rm impurity}}& = &\frac 2 {\pi a} \int dx   \left[ \eta_a(x) 
 \{\cos (\theta_{\rho -}- \phi_{\rho -})\cos (\theta_{\sigma -}
 -  \phi_{\sigma -}) \right. \nonumber \\ 
  &+& \left. \cos (\theta_{\rho -}+ \phi_{\rho -})
        \cos (\theta_{\sigma -}+ \phi_{\sigma -}) \} \right. \nonumber \\
 & & \left. + \xi_a(x) O_{CDW^\pi}(x)
                                                + {\rm H. c.}  -a \eta_s(x) \partial_x \phi_{\rho +} \right. \nonumber \\ 
 & & \left.+ \xi_s(x) O_{CDW^0}(x) 
                                                          +{\rm H. c.}  \right],
\end{eqnarray}
where 
 $O_{CDW^0}(x)$ ( $O_{CDW^\pi}(x)$) denotes a  operator  of 
  CDW (charge density wave) with in phase (out of phase) 
 between two chains and is defined by 
\begin{eqnarray}
 O_{CDW^0}(x)&=& e^{i \phi_{\rho+}} \{\sin \phi_{\rho -} 
     \cos \phi_{\sigma+} \cos \phi_{\sigma -} \nonumber \\
       &-&i \cos \phi_{\rho -} \sin \phi_{\sigma +} \sin \phi_{\sigma -}\}\label{eq:ocdwsym},  \\
 O_{CDW^\pi}(x)&=& e^{i \phi_{\rho+}} 
     \{ \cos \theta_{\rho-} \cos \theta_{\sigma -} \cos \phi_{\sigma +}\nonumber \\
    &+& i \sin \theta_{\rho-} \sin \theta_{\sigma -} \sin \phi_{\sigma +} \}. \label{eq:ocdwasym} 
      \end{eqnarray}
 The quantity 
 $ \Pi_{\nu,r}(x)$ is a variable 
 canonically conjugate to  $\phi_{\nu,r}(x)$ 
 and satisfies  
\begin{equation}
[ \phi_{\nu,r}(x),\Pi_{\nu',r'}(x') ]
      = i\delta(x-x')\delta_{\nu,\nu'}\delta_{r,r'} ,  
\end{equation}
 where 
\begin{equation}
\theta_{\nu,r}(x)=\pi \int_{-\infty}^x dx' \Pi_{\nu,r}(x') . 
\end{equation}
From bosonization rules, these quantities  are written as  $(m=0,\pi)$
\begin{eqnarray}
  \frac{c_{n,m,\sigma}}{\sqrt{a}}
   &=& e^{i k_F x} \frac{e^{i(\theta_{m,\sigma}-\phi_{m,\sigma})}}{\sqrt{2\pi a}}
  +e^{-i k_F x} \frac{e^{i(\theta_{m,\sigma}+\phi_{m,\sigma})}}{\sqrt{2\pi a}},
                          \\
 \phi_{\rho+}&=&\frac 1 2 (\phi_{0,\uparrow}+\phi_{0,\downarrow}
      +\phi_{\pi,\uparrow}+\phi_{\pi,\downarrow}), \\
\phi_{\rho-}&=&\frac 1 2 (\phi_{0,\uparrow}+\phi_{0,\downarrow}
             -\phi_{\pi,\uparrow}-\phi_{\pi,\downarrow}), \\
\phi_{\sigma+}&=&\frac 1 2 (\phi_{0,\uparrow}-\phi_{0,\downarrow}
              +\phi_{\pi,\uparrow}-\phi_{\pi,\downarrow}), \\
\phi_{\sigma-}&=&\frac 1 2 (\phi_{0,\uparrow}-\phi_{0,\downarrow}
             -\phi_{\pi,\uparrow}+\phi_{\pi,\downarrow}), \\
\Pi_{\rho+}&=&\frac 1 2 (\Pi_{0,\uparrow}+\Pi_{0,\downarrow}
        + \Pi_{\pi,\uparrow}+\Pi_{\pi,\downarrow}), \\
\Pi_{\rho-}&=&\frac 1 2 (\Pi_{0,\uparrow}+\Pi_{0,\downarrow}
         -\Pi_{\pi,\uparrow}-\Pi_{\pi,\downarrow}), \\
\Pi_{\sigma+}&=&\frac 1 2 (\Pi_{0,\uparrow}-\Pi_{0,\downarrow}
             +\Pi_{\pi,\uparrow}-\Pi_{\pi,\downarrow}), \\
\Pi_{\sigma-}&=&\frac 1 2 (\Pi_{0,\uparrow}-\Pi_{0,\downarrow}-\Pi_{\pi,\uparrow}
      +\Pi_{\pi,\downarrow}) .
\end{eqnarray}

The coupling constants of the bosonized Hamiltonian
(\ref{eq:bosonized_hamiltonian}) can be expressed as the function of those 
of the lattice Hamiltonian (\ref{eq:lattice_hamiltonian}). They are given by 
\begin{eqnarray}
                           \label{init_y}
 y_1&=&y_3=y_4=y_9=y_{10}=-y_5=-y_6=-y_8=\frac{Ua}{\pi v_F}  , 
                          \nonumber \\
 y_2&=&y_7=0, 
\end{eqnarray}
 and 
\begin{eqnarray}
K_{\nu -}&=&1, \\ 
 u_{\nu -}&=&v_F, \nonumber \\
 K_{\rho+}&=&\left(1+\frac{Ua}{\pi v_F}\right)^{-1/2}, \\
 u_{\rho +}&=& v_F \left(1+\frac{Ua}{\pi v_F}\right)^{1/2}, \\
  K_{\sigma+}&=&\left(1-\frac{Ua}{\pi v_F}\right)^{-1/2}, \\
  u_{\sigma+}&=&\left(1-\frac{Ua}{\pi v_F}\right)^{1/2}, \\
 {\cal D}_r&=&\frac{D^b_r a}{\pi v_F^2},
\end{eqnarray}
where
  $ D^f_r = D^b_r  = 2W\equiv D$,  
$ \overline{\eta_{r}(x)\eta_{r'}(x')} 
      = 2 D^f_r \delta(x-x') \delta_{r,r'}$ and  
$ \overline{\xi_{r}(x)\xi^*_{r'}(x')} 
  = 2 D^b_r \delta(x-x') \delta_{r,r'}$.
 $v_F$ is  
 the Fermi velocity in the absence of interaction. 
 We note that there is a freedom of sign 
  in choosing the initial condition\cite{schulz_moriond} although the result remains the same. 
 Actually,  the initial value of 
 $y_1$, $y_5$ and $y_{10}$ 
 in Eq.(\ref{init_y}) has a sign  opposite to 
   that  of \cite{tsuchiizu_confinement_ladder_long}, i.e, 
 the correspondence is given by the replacement 
 of  $\phi_{\sigma+} \rightarrow \phi_{\sigma+} + \pi/2$. 
We consider only  the backward scattering for impurity,  
 which is relevant to the localization.

\section{RG equations}\label{sec:renorm}

The RG equations can be derived from the Hamiltonian
(\ref{eq:bosonized_hamiltonian}) using the Operator Product Expansion
(OPE) approach\cite{cardy_scaling}. In the RGE, we neglect the velocity
differences between the different sectors. To be consistent with such an 
approximation, we also have to neglect $y_4,y_8$ in the RGE
Failing to do so leads to unphysical results such as  
a spin gap for $t_\perp=0$ and no
disorder  in disagreement with the exact results on the single
chain Hubbard model\cite{schulz_houches_revue}. 
It is also necessary to
expand the RG equations of $K_{\sigma \pm}$ to first order to ensure
SU(2) symmetry.  
The renormalization group equations read:
 \begin{eqnarray}\label{eq:full_rge}
 \frac{dK_{\rho+}}{dl} & = &-\frac{K_{\rho+}^2}{4}({\cal D}_s+{\cal D}_a) , \\
 \frac{dK_{\rho-}}{dl} & = &\frac 1 8 \left[y_1^2 + y_2^2 + y_3^2  
    + 2 {\cal D}_a  \right. \nonumber \\
    & &   \left.   -K_{\rho-}^2
     \left\{(y_5^2+y_6^2+y_7^2)J_0(4K_{\rho-} \tilde{t}_\perp) 
           +2 {\cal D}_s \right\}\right] , \\
 \frac{dy_{\sigma-}}{dl} 
     & = &\frac 1 4 \left[y_2^2 +y_6^2 J_0(4K_{\rho-} \tilde{t}_\perp) +y_{10}^2 
     +2 {\cal D}_a \right.  \nonumber \\
  & &  \left. -\left\{y_3^2+y_7^2 J_0(4K_{\rho-} \tilde{t}_\perp)+y_9^2
                        +2{\cal D}_s\right\}\right] ,  \\ 
 \frac{dy_{\sigma+}}{dl} 
    & = &-\frac 1 4 \left[y_1^2 +y_5^2 J_0(4K_{\rho-}
\tilde{t}_\perp)\right. \nonumber \\ &+& \left. y_9^2
     + y_{10}^2 +2 ({\cal D}_s+{\cal D}_a)\right] , \\
 \frac{dy_1}{dl} 
    & = &\left(1-\frac 1 {K_{\rho-}}-\frac{y_{\sigma +}}{2}\right)y_1
\nonumber \\ 
      &-&\frac 1 2 (y_2 y_{10}+y_3 y_9)-{\cal D}_a , \\
 \frac{dy_2}{dl} 
    & = &\left(1-\frac 1 {K_{\rho-}}+\frac{y_{\sigma -}}{2}\right)y_2
              -\frac 1 2 y_1 y_{10}-{\cal D}_a , \\ 
 \frac{dy_3}{dl} 
    & = &\left(1-\frac 1 {K_{\rho-}}-\frac{y_{\sigma -}}{2}\right)y_3 
                     -\frac 1 2 y_1 y_9 ,\\
 \frac{dy_5}{dl} 
   & = & \left(1-K_{\rho-}-\frac{y_{\sigma +}}{2}\right)y_5 \nonumber \\
              &-& \frac 1 2 ( y_7 y_9 + y_6 y_{10})+{\cal D}_s ,\\
 \frac{dy_6}{dl} 
   & = &\left(1-K_{\rho-}+\frac{y_{\sigma -}}{2}\right)y_6 
                 -\frac 1 2  y_5 y_{10} , \\
 \frac{dy_7}{dl} 
    & = &\left(1-K_{\rho-}-\frac{y_{\sigma -}}{2}\right)y_7
         -\frac 1 2  y_5 y_9  + {\cal D}_s , \\
 \frac{dy_9}{dl} 
     & = &-\frac 1 2 (y_{\sigma-} y_9 + y_{\sigma+} y_9 + y_1 y_3
\nonumber \\  
          &+& y_5 y_7 J_0(4K_{\rho-} \tilde{t}_\perp)) - {\cal D}_s , \\ 
 \frac{dy_{10}}{dl} 
     & = &-\frac 1 2 (y_{\sigma+} y_{10}-y_{\sigma-}y_{10} + y_1 y_2
\nonumber \\
         &+& y_5 y_6 J_0(4K_{\rho-} \tilde{t}_\perp)))-{\cal D}_a , \\
 \frac{d{\tilde{t}}_\perp}{dl} 
     & = &\tilde{t}_\perp 
      -\frac 1 {16} (y_5^2+ y_6^2+y _7^2) J_1(4K_{\rho-} \tilde{t}_\perp)) , \\
 \frac{d{\cal D}_a}{dl} 
    & = &\left\{2 - \frac 1 2 (K_{\rho+}+\frac 1 {K_{\rho-}}) 
         -\frac 1 4 y_{\sigma+}+\frac 1 4 y_{\sigma -}
\right. \nonumber \\ 
         &-& \left. \frac 1 2 y_1 -\frac 1 2 y_2 -\frac 1 2 y_{10}  \right\} {\cal D}_a ,
         \\
 \frac{d{\cal D}_s}{dl} 
    & = &\left\{2 - \frac 1 2 (K_{\rho+}+K_{\rho-}) 
         -\frac 1 4 y_{\sigma+}-\frac 1 4 y_{\sigma -} \right. \nonumber \\ 
        & +&\left. \frac 1 2 y_5 J_0(4K_{\rho-} \tilde{t}_\perp)   +\frac 1 2 y_7 J_0(4K_{\rho-} \tilde{t}_\perp) 
         -\frac 1 2 y_9  \right\} {\cal D}_s ,
                                 \label{eq:full_rgeend} \nonumber \\                              
\end{eqnarray}
where we have introduced $y_{\sigma\pm}$ and $\tilde{t}_\perp$ 
defined respectively  by $K_{\sigma
\pm}=1+y_{\sigma\pm}/2$ and $\tilde{t}_\perp=\frac{t_\perp a}{\pi
v_F}$.  

In these RG equations, we have neglected 
 the generation of a $4k_F$
  disorder from  $2k_F$ disorder   
 since the generation of such term in RG demands special treatment
\cite{fujimoto_mott+disorder_1ch}. 
In the present calculation, we will be finding 
SC$^d$ and SC$^s$ phases, in which  the $2k_F$ disorder is 
 regarded  as irrelevant.
However, based on the study of Ref.~\cite{orignac_2chain_long} we
already know that $4k_F$ disorder is always relevant in the SC$^d$ phase 
and relevant in the  SC$^s$ except for $K_{\rho+}>3/2$. 
Therefore, in reality these  phases of SC$^d$ and SC$^s$ 
  will be replaced by a pinned $4k_F$
CDW as discussed in Ref. \cite{fujimoto_mott+disorder_2ch}.  

\subsection{The limit of ${\cal D}=0$: the pure two-chain ladder}

In the limit where ${\cal D}=0$, a simplified form of the RGEs is
obtained as: 

\begin{eqnarray}\label{eq:clean_case}
\frac{dK_{\rho+}}{dl} & = & 0 \\ 
\frac{dK_{\rho-}}{dl} 
  & = &\frac 1 8 \left[y_1^2 + y_2^2 
    + y_3^2 \right. \nonumber \\ &-& \left.K_{\rho-}^2\left\{(y_5^2+y_6^2
    +y_7^2)J_0(4K_{\rho-} \tilde{t}_\perp) \right\}\right] , \\
\frac{dy_{\sigma-}}{dl} 
  & = &\frac 1 4 \left[y_2^2 +y_6^2 J_0(4K_{\rho-} \tilde{t}_\perp) 
     + y_{10}^2 \right. \nonumber \\  &-& \left. \left\{y_3^2+y_7^2 J_0(4K_{\rho-} \tilde{t}_\perp)
     +y_9^2\right\}\right] ,  \nonumber \\ \\
\frac{dy_{\sigma+}}{dl} 
   & = &-\frac 1 4 \left[y_1^2 +y_5^2 J_0(4K_{\rho-} \tilde{t}_\perp)
     +y_9^2+ y_{10}^2\right] , \\
\frac{dy_1}{dl} 
   & = &\left(1-\frac 1 {K_{\rho-}}-\frac{y_{\sigma +}}{2}\right)y_1 
      -\frac 1 2 (y_2 y_{10} + y_3 y_9) , \\
\frac{dy_2}{dl} 
   & = &\left(1-\frac 1 {K_{\rho-}}+\frac{y_{\sigma -}}{2}\right)y_2
       -\frac 1 2 y_1 y_{10} , \\ 
\frac{dy_3}{dl} 
         & = &\left(1-\frac 1 {K_{\rho-}}-\frac{y_{\sigma -}}{2}\right)y_3 
       -\frac 1 2 y_1 y_9 , \\
\frac{dy_5}{dl} 
      & = & \left(1-K_{\rho-}-\frac{y_{\sigma +}}{2}\right)y_5
        -\frac 1 2 ( y_7 y_9 + y_6 y_{10}) , \\
\frac{dy_6}{dl} 
      & = &\left(1-K_{\rho-}+\frac{y_{\sigma -}}{2}\right)y_6 
        -\frac 1 2  y_5 y_{10} , \\
\frac{dy_7}{dl} 
     & = &\left(1-K_{\rho-}-\frac{y_{\sigma -}}{2}\right)y_7
       -\frac 1 2 y_5 y_9  , \\
\frac{dy_9}{dl} 
     & = &-\frac 1 2 (y_{\sigma-} y_9 + y_{\sigma+} y_9 + y_1 y_3
\nonumber \\ 
      & +& y_5 y_7 J_0(4K_{\rho-} \tilde{t}_\perp)) , \\ 
\frac{dy_{10}}{dl} 
       & = &-\frac 1 2 (y_{\sigma+} y_{10}-y_{\sigma-}y_{10} 
       + y_1 y_2 \nonumber \\  &+& y_5 y_6 J_0(4K_{\rho-} \tilde{t}_\perp))) , \\
\frac{d{\tilde{t}}_\perp}{dl} 
     & = &\tilde{t}_\perp 
     -\frac 1 {16} ( y_5^2+ y_6^2+y _7^2) J_1(4K_{\rho-} \tilde{t}_\perp)) . 
\end{eqnarray}

These  equations are identical to those previously derived in
\cite{tsuchiizu_confinement_ladder_long}.
Taking the limit of $t_\perp \to +\infty$, which is the limit of the two 
chain system, these equations are further
simplified\cite{yoshioka_coupledchains_interaction,yoshioka_2ch}: 

\begin{eqnarray}\label{eq:two_chain_regime}
\frac{dK_{\rho-}}{dl} 
  & = &\frac 1 8( y_1^2 + y_2^2 + y_3^2 ) , \\
      \frac{dy_{\sigma-}}{dl} 
  & = &\frac 1 4 \left[y_2^2 + y_{10}^2 -( y_3^2 + y_9^2 )\right] , \\
\frac{dy_{\sigma+}}{dl} 
      & = &- \frac 1 4 \left[y_1^2 + y_9^2+ y_{10}^2\right] , \\
\frac{dy_1}{dl} 
     & = &\left(1-\frac 1 {K_{\rho-}}-\frac{y_{\sigma +}}{2}\right)y_1 
         -\frac 1 2 (y_2 y_{10} + y_3 y_9) , \\
\frac{dy_2}{dl} 
        & = &\left(1-\frac 1 {K_{\rho-}}+\frac{y_{\sigma -}}{2}\right)y_2
          -\frac 1 2 y_1 y_{10} , \\ 
\frac{dy_3}{dl} 
       & = &\left(1-\frac 1 {K_{\rho-}}-\frac{y_{\sigma -}}{2}\right)y_3 
        -\frac 1 2 y_1 y_9 , \\
\frac{dy_9}{dl} 
        & = &-\frac 1 2 (y_{\sigma-} y_9 + y_{\sigma+} y_9 + y_1 y_3 ) , \\ 
\frac{dy_{10}}{dl} 
       & = &-\frac 1 2 (y_{\sigma+} y_{10}- y_{\sigma-} y_{10} + y_1 y_2 ) . 
\end{eqnarray}

Obviously, one has: $K_{\rho-}\to +\infty, y_{\sigma+}\to -\infty$. 
Moreover, it is easily seen from initial conditions that originally
$|y_2|<|y_3|$ and $|y_9|=|y_{10}|$ , resulting in
$\frac{dy_{\sigma-}}{dl}<0$. Thus, for small $l$, one should have
$y_{\sigma-}<0$. This in turn makes $y_2,y_{10}$  less relevant and
results in both $\frac{dy_{\sigma-}}{dl}$ and $y_{\sigma-}$ becoming more
negative. As a result, one finds $y_{\sigma-} \to -\infty$. Thus
$\theta_{\rho-}$ and 
 $\phi_{\sigma\pm}$ become locked in the ladder system. 
Further analysis of these RGE shows that there are two possible fixed
points\cite{schulz_moriond,schulz_2chains} depending on the sign of $U$ 
 as represented on table \ref{tab:fixed-points}. 
The  fixed point for $U>0$ correspond to the formation 
of d-wave superconductivity, whereas the one for $U<0$
 corresponds to the formation of s-wave superconductivity.

\subsection{The case $t_\perp=0$: single chain limit}
This case corresponds to two decoupled disordered Hubbard chains and has 
been analyzed in details in Ref.\cite{giamarchi_loc}. A crucial test of
the validity of our RG equations is whether we can recover the RG
equations of Ref. \cite{giamarchi_loc} by putting $t_\perp=0$ in
Eqs. (\ref{eq:full_rge})-(\ref{eq:full_rgeend}).  
For $t_\perp=0$, it is straightforward to show that the solution of the
RG equations (\ref{eq:full_rge})-(\ref{eq:full_rgeend}) satisfies:
\begin{eqnarray}\label{eq:1chain_conditions}
 {\cal D}_s(l) &=&{\cal D}_a(l)={\cal D}(l), \\
  y_6(l) & = & -y_3(l), \\
  y_7(l) & = &-y_2(l), \\
 K_{\rho-}(l) & = & K_{\sigma -}(l)=1, \\
 y_{\sigma+}(l) & = & y_1(l)=y_9(l)=y_{10}(l)\nonumber \\ 
        & = & -y_5(l)=y_2(l)+y_3(l)=y(l). 
\end{eqnarray} 
As a result, the RG equations~(\ref{eq:full_rge})-(\ref{eq:full_rgeend})
 are  reduced  to the
following equivalent set:
\begin{eqnarray}
 \frac{dK_{\rho+}}{dl} & = & -\frac{K_{\rho+}^2}2 {\cal D} , \\
 \frac{dy}{dl} & = & -y^2-{\cal D} , \\
 \frac{dy_2}{dl} & = & -\frac{y^2}2 -{\cal D} , \\
 \frac{dy_3}{dl} & = & -\frac{y^2}2 , \\
 \frac{d{\cal D}}{dl} & = & \left(\frac 3 2 - \frac{K_{\rho+}}{2} 
           - \frac 5 4 y - \frac 1 2 y_2 \right) {\cal D} .
\end{eqnarray} 

Using the expansion: $K_{\rho+}=1+{y_{\rho+}}/{2}$, we rewrite
 the RG equation for $K_{\rho+}$ in the form:
\begin{equation}
\frac{dy_{\rho+}}{dl}=-{\cal D} ,
\end{equation}
with $y_{\rho+}(0)=-y(0)$. It is then easily seen that
$y_2(l)=(y_{\rho+}(l)+y(l))/2$, 
 for all $l$. 
This allows to rewrite the RGE for ${\cal D}$ as:
\begin{eqnarray}
\frac 1 {{\cal D}} \frac{d{\cal D}}{dl}
       =  2 -K_{\rho+}-\frac 3 2 y_{\sigma +} .
\end{eqnarray} 
The RGE equations are thus reduced  to those of  the single chain
case (see Eq. (3.12) in Ref. \cite{giamarchi_loc}). Then, our RG
equations have the correct limit of $t_\perp=0$. 
Obviously, for ${\cal D}=0$, we recover the
RGEs that can be applied
  to a single chain system. In particular, we can easily
check that the system develops a spin gap for $U<0$ but remains gapless
for $U>0$. Thus, we see that 
 the RGEs (\ref{eq:full_rge})-(\ref{eq:full_rgeend}) correctly
capture the limit of a single chain system.

\section{The limit $t_\perp \to \infty$: two chain regime}
In this section, we  consider the limit $t_\perp \to \infty$ of 
 the equations (\ref{eq:full_rge})-(\ref{eq:full_rgeend})
  in order to make contact with the work 
of
Refs.\cite{fujimoto_mott+disorder_2ch,sandler_disorder_qwires,mori_ladders_loc,mori_persistent_2ch}.
In the limit $t_\perp \to \infty$, the Bessel functions are negligible,
so that Eq. (\ref{eq:full_rge})-(\ref{eq:full_rgeend}) are reduced to:

\begin{eqnarray}
 \frac{dK_{\rho+}}{dl} & = &-\frac{K_{\rho+}^2}{4}({\cal D}_s+{\cal D}_a) , \\
 \frac{dK_{\rho-}}{dl} & = &\frac 1 8 \left[y_1^2 + y_2^2 + y_3^2  
    + 2 {\cal D}_a  \right. \nonumber \\
    & &   \left.   -2K_{\rho-}^2 {\cal D}_s \right] , \label{eq:krm2ch}\\
 \frac{dy_{\sigma-}}{dl} 
     & = &\frac 1 4 \left[y_2^2  +y_{10}^2 
     +2 {\cal D}_a \right.  \nonumber \\
  & &  \left. -\left\{y_3^2 + y_9^2
                        +2{\cal D}_s\right\}\right] ,\label{eq:ysm2ch}  \\ 
 \frac{dy_{\sigma+}}{dl} 
    & = &-\frac 1 4 \left[y_1^2 +y_9^2
     + y_{10}^2 +2 ({\cal D}_s+{\cal D}_a)\right] , \label{eq:ysp2ch}\\
 \frac{dy_1}{dl} 
    & = &\left(1-\frac 1 {K_{\rho-}}-\frac{y_{\sigma +}}{2}\right)y_1
\nonumber \\
     & -& \frac 1 2 (y_2 y_{10}+y_3 y_9)-{\cal D}_a , \label{eq:y1-2ch}\\
 \frac{dy_2}{dl} 
    & = &\left(1-\frac 1 {K_{\rho-}}+\frac{y_{\sigma -}}{2}\right)y_2
              -\frac 1 2 y_1 y_{10}-{\cal D}_a, \label{eq:y2-2ch}\\ 
 \frac{dy_3}{dl} 
    & = &\left(1-\frac 1 {K_{\rho-}}-\frac{y_{\sigma -}}{2}\right)y_3 
                     -\frac 1 2 y_1 y_9 ,\label{eq:y3-2ch}\\
 \frac{dy_9}{dl} 
     & = &-\frac 1 2 (y_{\sigma-} y_9 + y_{\sigma+} y_9 + y_1 y_3  
           - {\cal D}_s ,\label{eq:y9-2ch} \\ 
 \frac{dy_{10}}{dl} 
     & = &-\frac 1 2 (y_{\sigma+} y_{10}-y_{\sigma-}y_{10} + y_1 y_2 
         -{\cal D}_a , \label{eq:y10-2ch} \\
 \frac{d{\cal D}_a}{dl} 
    & = &\left\{2 - \frac 1 2 (K_{\rho+}+\frac 1 {K_{\rho-}}) 
         -\frac 1 4 y_{\sigma+}+\frac 1 4 y_{\sigma -}  \right. \nonumber \\
        & -& \left. \frac 1 2 y_1 -\frac 1 2 y_2 -\frac 1 2 y_{10}  \right\} {\cal D}_a , \label{eq:Da2ch}
                         \\
 \frac{d{\cal D}_s}{dl} 
    & = &\left\{2 - \frac 1 2 (K_{\rho+}+K_{\rho-}) 
         -\frac 1 4 y_{\sigma+}-\frac 1 4 y_{\sigma -}  
         -\frac 1 2 y_9  \right\} {\cal D}_s  \label{eq:Ds2ch}
                                \nonumber , \\
\end{eqnarray}

Let us compare our equations with those in \cite{mori_persistent_2ch}.
Equations  (11), (12), (14), (15) in Ref. \cite{mori_persistent_2ch}
correspond
respectively to
equations  (\ref{eq:Da2ch}), (\ref{eq:y9-2ch}),
(\ref{eq:y10-2ch}) and (\ref{eq:y1-2ch}) in the present  paper.
However, in     
 Eqs. (12), (14) and (15) of  
 \cite{mori_persistent_2ch} 
 we do not find  terms arising from  coupling between interactions, 
which give rise to the same order. 
More importantly,
Eq. (\ref{eq:Ds2ch}) is in disagreement with Eq. (10) in
Ref. \cite{mori_persistent_2ch} . 
The reason is that  in
Ref. \cite{mori_persistent_2ch}, the $y_5$ term was kept although it
should drop from the Hamiltonian when $t_\perp$ becomes sufficiently
large. 
The correct initial RG equations for weak disorder and large $t_\perp$
thus read:
\begin{eqnarray}
\frac{d{\cal D}_a}{dl}&=&\left(1-\frac{Ua}{\pi v_F}\right){\cal D}_a \label{eq:Da_initial_rg}\\
\frac{d{\cal D}_s}{dl}&=&\left(1-\frac{Ua}{2\pi v_F}\right){\cal D}_s \label{eq:Ds_initial_rg}
\end{eqnarray}
These equations imply that antisymmetric disorder is more relevant
than symmetric disorder in the case of attractive interaction and less
relevant for repulsive interactions. Such result can be understood in
the following way: for repulsive interactions, the system tends to
form interchain Cooper pairs and thus to have a repartition of charges
that is symmetric between the two ladders. This makes the system less
sensitive to antisymmetric disorder. For attractive interactions, the
system prefers to form intrachain Cooper pairs. If electrons of the
Cooper pair can tunnel on the opposite chain in a virtual process, the
energy of the Cooper pair is lowered. This results in interchain
repulsion of the pairs and thus antisymmetric CDW fluctuations. This
mechanism obviously makes antisymmetric disorder a more relevant
perturbation.
 
Comparing with Ref. \cite{fujimoto_mott+disorder_2ch}, we have similar
equations 
once we put the Umklapp 
process of Ref. \cite{fujimoto_mott+disorder_2ch} equal to
zero. However, an important difference between the present paper and
\cite{fujimoto_mott+disorder_2ch} is that in
\cite{fujimoto_mott+disorder_2ch} a random $t_\perp$ term is also
included in the Hamiltonian. This random $t_\perp$ can result in closure
of the transverse charge gap and thus can lead to 
a rather different physics from the one discussed here, closer to the
one considered in
\cite{mori_ladders_loc,mori_persistent_2ch,sandler_disorder_qwires}.
We will compare in more details our results and those of
\cite{fujimoto_mott+disorder_2ch} in the next section.

\section{Results}\label{sec:results}

From  
RG equations (\ref{eq:full_rge})-(\ref{eq:full_rgeend}), 
 we can distinguish  
four different phases 
 where the corresponding fixed points are shown on table 2 
( explained  below).

\subsection{SCd Phase}\label{sec:scd_results}
\begin{figure}

\epsfig{file=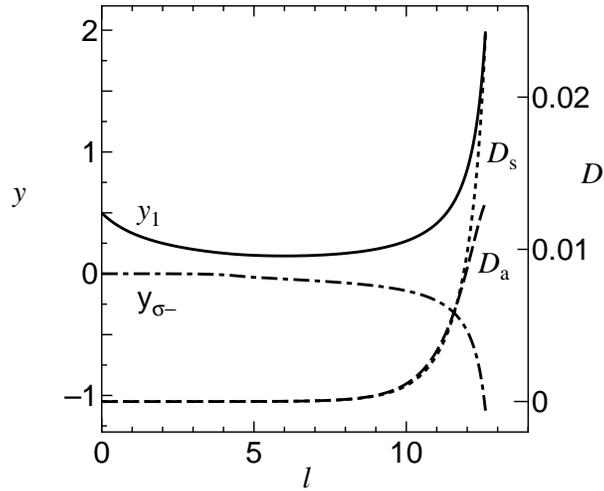,angle=0,width=8cm}
\caption{
RG flow as a function of $l$ for the SCd phase 
with  $Ua/\pi v_{\rm F} =0.5$, $t_{\perp}= 10^{-2}$ and $D = 10^{-7}$ 
 where the left axis is for $y_1$  and $y_{\sigma -}$ 
  and the right axis is for $D_{\rm a}$ (dashed curve) and $D_{\rm s}$ (dotted curve).   
}
\label{fig:scd}

\end{figure}

For large repulsive interaction and small disorder,  
 we obtain  the SCd phase, where 
 strong coupling in interaction  is reached before strong
coupling in disorder. The resulting flow of coupling constants is
represented on Fig. \ref{fig:scd}.
 The ground state has $\langle \theta_{\rho -} \rangle=0$, $\langle\phi_{\sigma
-}\rangle=\langle\phi_{\sigma +}\rangle=\frac \pi 2$ 
 (I in table 2).
This phase to the approximation of RG is a metallic phase. It has been
discussed in \cite{fujimoto_mott+disorder_2ch} as case A. However,
including coupling to $4k_F$ disorder, it is possible to show that such
 a phase is in  fact localized  
\cite{orignac_2chain_short,orignac_2chain_long,fujimoto_mott+disorder_2ch} 
due to pinning of 
 $4k_F$ CDW fluctuations by disorder \cite{orignac_2chain_short,orignac_2chain_long}. 
We can see that in this phase, $D_s$ is more relevant than $D_a$ which
is consistent with
Eqs. (\ref{eq:Da_initial_rg})--(\ref{eq:Ds_initial_rg}). 
For stronger interaction, this behavior can also be understood from
Eqs. (\ref{eq:ocdwsym})--(\ref{eq:ocdwasym}).  Replacing $\theta_{\rho-},\phi_{\sigma \pm}$ by
their expectation values in these equations, 
 we obtain $O_{CDW^0} \sim e^{i \phi_{\rho+}}
\cos \phi_{\rho-}$ and $O_{CDW^\pi} \sim  e^{i \phi_{\rho+}} (\cos
\theta_{\sigma-} \cos \phi_{\sigma+} + i \sin \theta_{\rho-} \sin
\theta_{\sigma-})$ where we have kept only the operators with
power-law or exponential decay. Clearly, $O_{CDW^\pi}$ contains more
disordered operators with exponential decay than
$O_{CDW^0}$. As a result, CDW$^0$ are dominating over the CDW$^\pi$ in
the SCd phase, in agreement with the picture of interchain
pairing. This also implies that the random potential that couples to
$O_{CDW^0}$ is more relevant than the one that coupled to $O_{CDW^\pi}$
even deep in the SCd phase.

\subsection{SCs phase}\label{sec:scs_results}
\begin{figure}

\epsfig{file=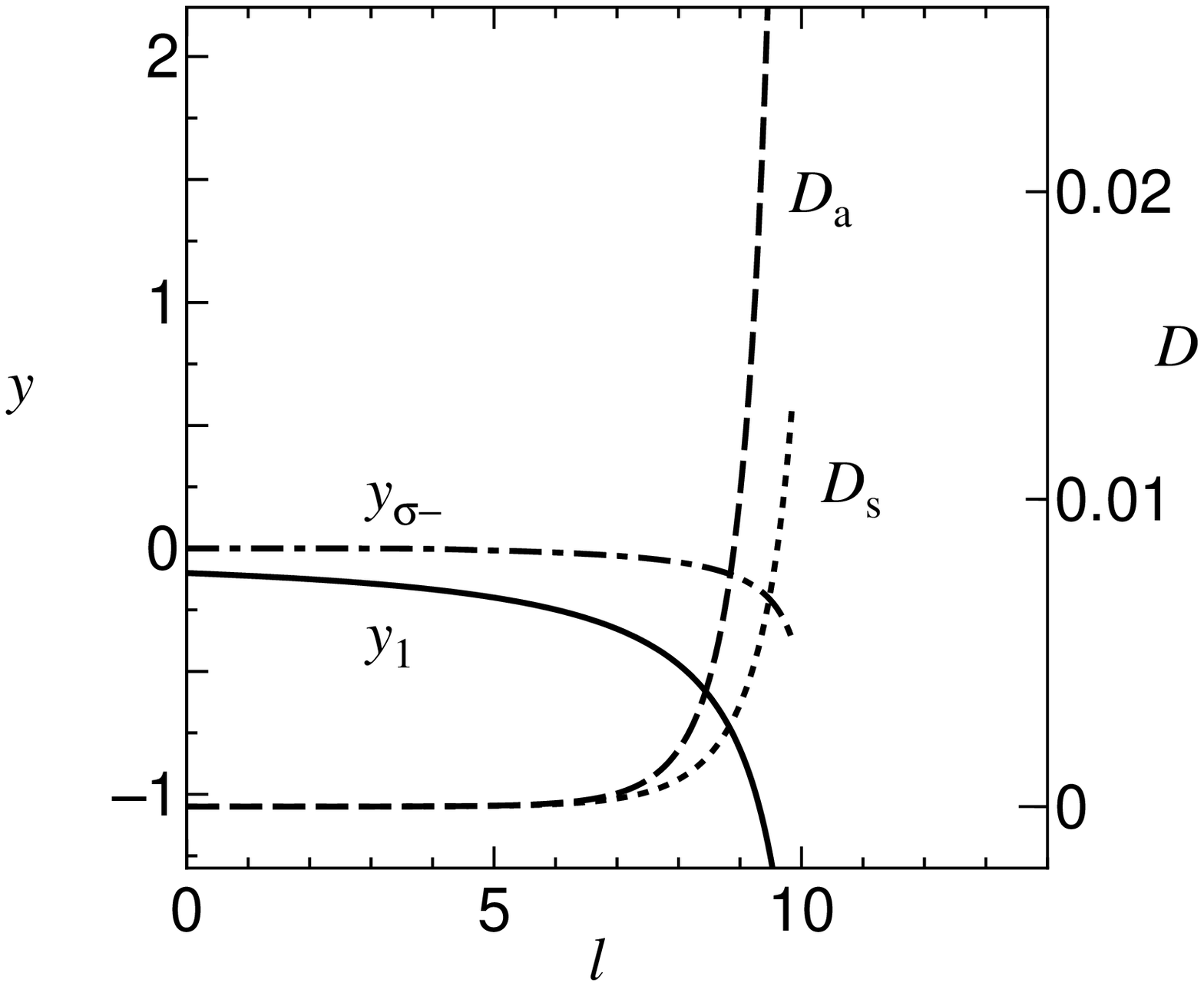,angle=0,width=8cm}
\caption{
RG flow as a function of $l$ for the SCs phase 
with  $Ua/\pi v_{\rm F} = -0.1$, $t_{\perp}= 10^{-2}$ and $D = 10^{-7}$ 
 where the notations are the same as Fig. 1.
}
\label{fig:scs}

\end{figure}

For interaction being moderately attractive  and disorder being small, 
 we obtain the phase  where 
 strong coupling in interaction is reached  before strong
coupling in disorder. The resulting flow of coupling constants is
represented on Fig. \ref{fig:scs}. 
In this case, the ground state has $\langle \theta_{\rho -}\rangle=\langle\phi_{\sigma
-}\rangle=\langle\phi_{\sigma +}\rangle=0$ 
 (IIb in table 2). 
Similarly to the SCd phase, the SCs phase is localized by disorder due
to the presence  of subdominant $4k_F$
fluctuations~\cite{orignac_2chain_short,orignac_2chain_long}. However,
for attractive enough interactions
\cite{orignac_2chain_short,orignac_2chain_long} the $4k_F$ CDW can be
depinned by quantum fluctuations in contrast with the SCd phase. 
It can be seen that in the SCs phase, $D_a$ is increasing faster than
$D_s$ in agreement with
Eqs. (\ref{eq:Da_initial_rg})--~(\ref{eq:Ds_initial_rg}). 
The persistence of this behavior even when the spin gap is well formed
can be understood by the same arguments as in \ref{sec:scd_results}.
 Comparing Eqs. (\ref{eq:ocdwsym}) and
(\ref{eq:ocdwasym}). In Eq. (\ref{eq:ocdwasym}), only the operator $\cos 
\theta_{\sigma-}$ has exponentially decaying correlation, whereas in
Eq. (\ref{eq:ocdwsym}) the three operators $\cos \phi_{\rho-}$, $\sin
\phi_{\sigma-}$ and $\sin \phi_{\sigma+}$ are disordered with
exponentially decaying correlations. 
As a result, the $CDW^0$ have a much faster exponential decay than the
$CDW^\pi$ fluctuations inside the SCs phase. In turn, this makes $D_s$
much less relevant than $D_a$. Physically, this can be
understood by noting that attractive interactions form Cooper pair
inside a single chain, leading to subdominant $CDW^\pi$ fluctuations 
whose   transverse charge fluctuation is compatible with that of 
SCs fluctuations.

\subsection{PCDW$^\pi$ phase}\label{sec:pcdwpi_results}
\begin{figure}

\epsfig{file=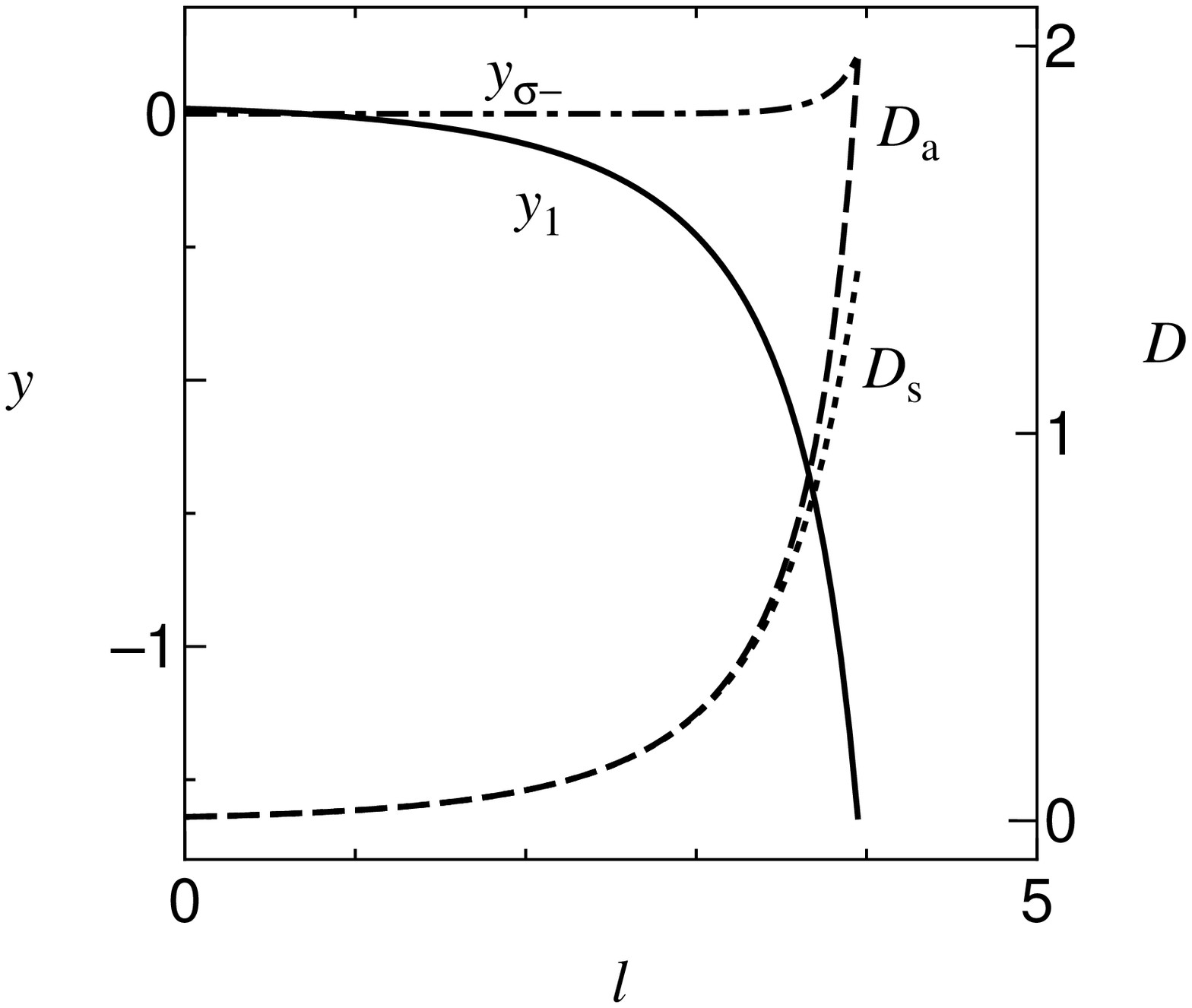,angle=0,width=8cm}
\caption{
RG flow as a function of $l$ for the  the PCDW phase 
with  $Ua/\pi v_{\rm F} = 0.01 $, $t_{\perp}= 10^{-2}$ and $D = 10^{-3}$ 
 where the notations are the same as Fig. 1.
}
\label{fig:pcdw}

\end{figure}

For intermediate  disorder, we obtain 
 the PCDW$^\pi$ phase
 where  strong coupling in the RG equations is obtained for
disorder first. The flow of the coupling constants is represented on
Fig. \ref{fig:pcdw}.  In this phase, 
one finds that $D_a$ is diverging faster 
than $D_s$. This corresponds to the  long range ordering with 
$\langle\theta_{\rho -}\rangle= \langle\phi_{\sigma +}\rangle
=\langle \theta_{\sigma -} \rangle=0$
 (III in table 2). 
This phase is the pinned $CDW^\pi$ of
Ref. \cite{orignac_2chain_long} and has also been discussed within a
RG approach in
\cite{fujimoto_mott+disorder_2ch}. 
We note that such a phase appears only for disorder strong enough to
compete with interaction 
 ( i.e.,  not in the limit of infinitesimal disorder)  
 for the present model. The presence of such phase in the ladder
system can have important consequences for transport properties since
pinning properties of the   $CDW^\pi$ are very different from the
$4k_F$ CDW. 

\subsection{Confinement}
\begin{figure}

\epsfig{file=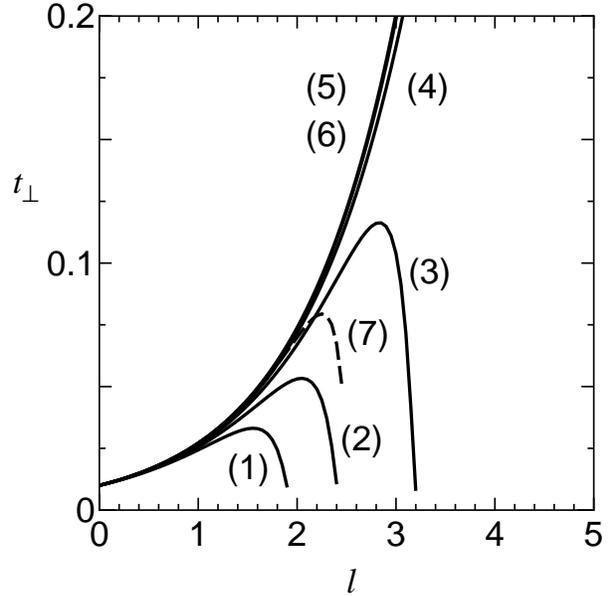,angle=0,width=8cm}
\caption{
RG flow of $t_{\perp}$ showing both 
confinement (irrelevant $t_{\perp}$) and deconfinement (relevant $t_{\perp}$) 
with  $Ua/\pi v_{\rm F}$ = -0.5(1), -0.4(2), -0.3(3), -0.2(4), -0.1(5) and  0(6) 
 where $t_{\perp} = 10^{-2}$ and $D = 10^{-3}$. 
 The curve (7) also shows the confinement for 
 $Ua/\pi v_{\rm F}$ = 0.1,  
  $t_{\perp} = 10^{-2}$ and $D = 10^{-1}$.
}
\label{fig:confine}

\end{figure}

If strong coupling in disorder is reached before $\tilde{t}_\perp(l)=1$, there
is a possibility of confinement of carriers in the transverse direction
 corresponding to the irrelevance of $\tilde{t}_\perp$
\cite{suzumura_tsuchiizu_gruner,tsuchiizu_confinement_ladder_long}
 (IV in table 2).  
This is represented on Fig.~\ref{fig:confine} 
 (curve (7)). Note that on the figures $t_\perp$ stands for
$\tilde{t}_\perp$.  
This confinement can be understood as the 
formation of a localized state at an energy scale higher than
$t_\perp$. 
 This time, a spin
gap is formed at an  energy scale higher than $t_\perp$. 
As a result, we have to distinguish two types of pinned $2k_F$ CDWs. 
One is the pinned $2k_F$ CDW$^\pi$ of 
 the two chain system (III in table 2), 
and the second one
is the combination of two pinned $2k_F$ CDW in each chain with irrelevant
interchain hopping (IV in table 2). 
 In the first case, there is a definite phase
relation between the CDW in chain 1 and the CDW in chain 2, while in the 
second case,  such relation is less definite. 
Such regime could not be obtained with the approximations used in
\cite{fujimoto_mott+disorder_2ch} or \cite{orignac_2chain_long} which
are valid only when $\tilde{t}_\perp$ is relevant. 
Further, we note that 
 confinement can also result from the presence of an
attractive interaction in the absence of disorder
 as seen  curves (1), (2) and  (3) in  Fig.~\ref{fig:confine})
 (IIa in table 2). 

 Here the difference in the disordered state  
between IIa and IV is examined besides the mechanism for confinement. 
In the region IIa of Fig. \ref{fig:pd},  a Josephson coupling is present and dominates 
the low energy properties. The physics of this region can be understood
by describing the Cooper pair as hard core bosons. 
The problem is  reduced to a disordered  bosonic ladder with interchain
hopping considered in Ref. \cite{orignac_2chain_bosonic}. 
 Thus it is found that the phase  
  in region IIa is  a pinned $4k_F$ CDW.
In region IV, the localization effects are stronger than the
Josephson coupling, leading to two 
 independently  pinned $2k_F$ CDWs.

\subsection{Criterion for transitions between the different phases}
We examine boundaries between 
 several types of phases ,which  are shown in Fig.~\ref{fig:pd}. 
\subsubsection{Confinement transition}\label{sec:confinement_transition}
A criterion for disorder confinement can be obtained by comparing the
length scale $l_{loc.}$  ( ${\cal D}(l_{loc.})=1$) 
 with the length scale
$l_{2ch.}$ ( $t_{\perp}(l_{2ch.})=1$). 
 Clearly, if $l_{loc.}<l_{2ch.}$
the system is localized before entering into the two-chain regime and thus
the ground state will be formed of two chains pinned on their own
disorder. For weak interactions, this criterion corresponds to ${\cal
D}(0)>t_{\perp}(0)$. As can be seen on fig. \ref{fig:pdtu}, it gives
the correct boundary between the pinned $2k_F$ CDW$^\pi$ (phase (III))
and the independently pinned  $2k_F$ CDWs (phase (IV)). 
 For stronger interactions, the criterion for
confinement is 
 $ E_{loc.} =   (v_{F}/a)e^{-l_{loc.}} = t_\perp$. Since the pinning energy is
enhanced in the presence of interactions, this implies that
interactions enhance confinements effects as can be seen on
Fig. \ref{fig:pdtu}. 
In the case of confinement by intrachain interactions one has to
compare the scale at which $y(l)=-\infty$ (for ${\cal D}=0$) to the scale at which
$\tilde{t}_\perp(l)\sim 1$. 
Since we have: 
\begin{equation}
y(l)=\frac{y(0)}{1+y(0)l} ,
\end{equation}
we find that confinement occurs when:
\begin{equation}
y(0)=\frac{1}{\ln (\tilde{t}_\perp(0))} .
\end{equation}
For $\tilde{t}_\perp=10^{-2}$, this gives confinement when 
 $U a /\pi v_{F} = -0.22$, which is the  correct value for
Fig.~\ref{fig:pd} when ${\cal D}<10^{-8}$. 

In Fig.~\ref{fig:pd}, we see that the two confinement regimes merge to give rise to 
a single confined phase. 
An important question is whether the transition between the confined and 
the deconfined phase is a true phase transition (i. e. whether one can
find an order parameter for the confined or deconfined phase) or if it
is only a crossover. 
 In the case of a commensurate potential ( at half-filling), it is well 
known
\cite{tsuchiizu_confinement_ladder_long,tsuchiizu_donohue_S_G,lehur_confinement} 
  that no order parameter
exists than can distinguish the confined from the deconfined
phase. Thus, we do not expect to see a true transition in the disordered 
phase. Instead, we expect a progressive loss of phase coherence between
the two CDWs in chain 1 and 2.
 
\subsubsection{Transition between PCDW$^\pi$ and SC phases}
This transition is obtained in the deconfined regime. A criterion for
this transition can be obtained by comparing the $2k_F$ CDW pinning
energy with the gap in $\sigma-$ sector. When the pinning energy is
above the gap, the system gains more energy by being in the PCDW phase. 
The transverse gap in $\sigma-$ is obtained by applying single chain RG
for $\tilde{t}_\perp e^l \ll 1$ and two chain RG for $\tilde{t}_\perp e^l \gg 1$.
Comparing the transverse gap with the pinning energy leads to a good
agreement with the phase boundary between (I) and (III).

\subsection{Global phase diagram} 
\begin{figure}

\epsfig{file=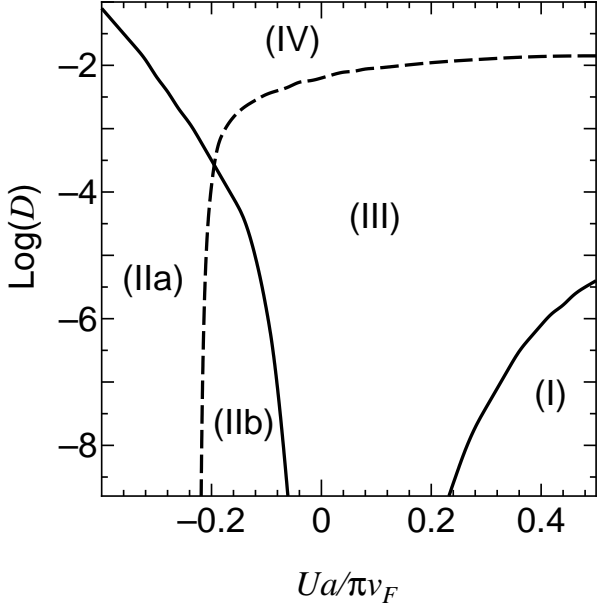,angle=0,width=8cm}
\caption{
Phase diagram in the $(\log({\cal D}),U)$ plane for a fixed
 $t_\perp=10^{-2}$. 
 The respective phases denote 
 the  SCd state ( region (I)), 
 the confined SCs state (region (IIa)),  the SCs state (region (IIb)) 
 the  pinned CDW$^{\pi}$ state (region (III)) and 
  the  pinned CDW$^{\pi}$ state with confinement (region (IV)). }  
\label{fig:pd}

\end{figure}

\begin{figure}[htbp]

\epsfig{file=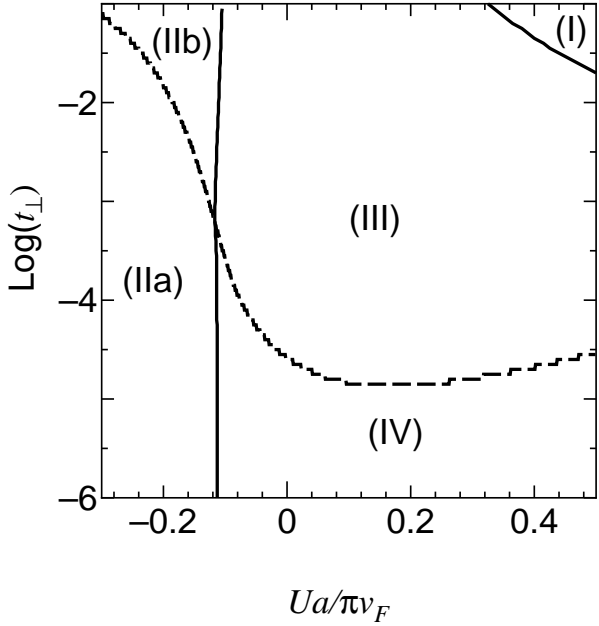,angle=0,width=8cm}
\caption{Phase diagram in the $(t_\perp,U)$ plane for a fixed
 $D=10^{-5}$ where the notations are the same as Fig. 5. }

\label{fig:pdtu}
\end{figure}

The global phase diagram is shown in Fig.~\ref{fig:pd} with the fixed 
 $t_{\perp} = 10^{-2}$. 
Based on the criterion in the previous subsection, 
 the boundary between pinned CDW phase (region (III)) 
and superconducting phases (region (I) or region (IIa)) 
 is given by 
\begin{eqnarray}
y_{\sigma-}(l_{loc.})=0 .
\end{eqnarray}
 The quantity  $l_{loc.}$ is the scale at which  one
of the coupling constants of disorder is equal to one.
This is also understood from Figs. \ref{fig:scd}, \ref{fig:scs}, and
\ref{fig:pcdw}  where
  the fixed point of $y_{\sigma-}$ is given by 
 $ - \infty$  for regions (I) and (IIa)
 and 
 $  \infty$  for regions (III). In the confined region, the
corresponding boundary is given by $y_{\sigma-}(\infty)=0$.   
With increasing $D$, the pinned  CDW region is enlarged.  
The region for SCs state is much larger than that for SCd state 
 since the spin gap exists even for the single chain. 
The boundary between region (I) and (III) is also obtained  from  
 the condition   $y_1(\infty)=0$ 
 since the fixed point for $y_1$ is given by   
  $ \infty$  for region (I) and  $ - \infty$  for region (III).
The boundary between (III) and (IIb) ( or between (IV) and (IIb)) 
  is also given by $y_2(\infty)/y_3(\infty) = 1$
 since $y_2(\infty) / y_3 (\infty) > 1 (<1)$ for region(III) and (IV) 
  (region (IIb) and (IIa)). 
The boundary between the confined and deconfined phase is given by the
condition 
\begin{eqnarray}
 \tilde{t}_\perp(l_c)=1. 
 \end{eqnarray}
 The quantity  $l_c$ denotes a value at which $t_{l}$ takes a maximum. 
The boundary between confinement and deconfinement 
in the pinned CDW region 
 moves continuously to that in th SCs state. 
 The former is determined by disorder 
 while  the latter is determined by attractive interaction. 
 Such a continuous variation of the boundary indicates a fact that 
 the confinement occurs by the combined effect of 
 interchain hopping, intrachain interaction and disorder. 

Another  global phase diagram is shown in Fig. \ref{fig:pdtu} 
 in   the $(t_{\perp},U)$ plane with a fixed 
 $D= 10^{-5}$ where the notations are the same as Fig.~\ref{fig:pd}.  
 The effect of interchain hopping on confinement is opposite to 
  that of disorder in Fig.~\ref{fig:pd}.    
The  $t_{\perp}$ dependence on the boundary between (III) and 
(IIa) ((III) and (IIb)) is small while it is noticeable 
 for the boundary between (III) and (IV) ( (III) and   (I)). 

\section{Discussion}\label{sec:discussion}

We have examined the effect of interchain hopping on  disordered 
 two-coupled 
chains using RG equations that enable us to describe the crossover from
the single chain to two chain regime.
Using the RG equations, we have shown that a $2k_F$ CDW$^\pi$ could be
obtained when disorder was strong enough compared to interaction.

\begin{figure}[htbp]

\epsfig{file=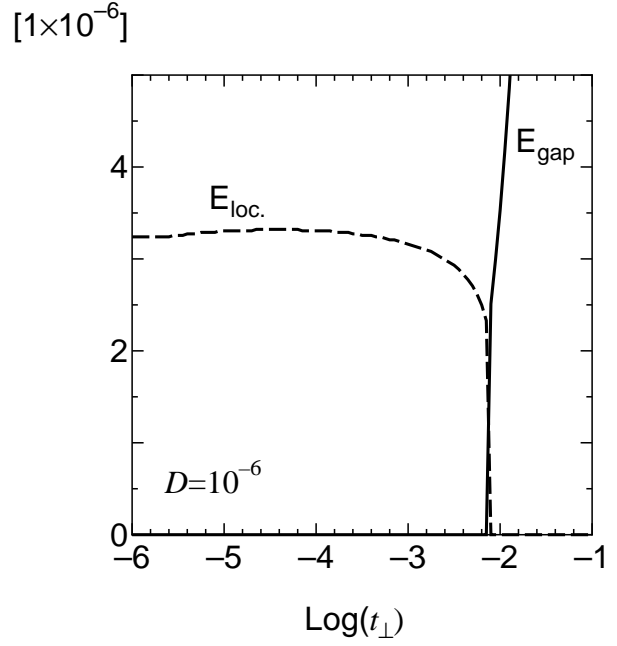,angle=0,width=8cm}
\caption{Behavior of $2k_F$ pinning energy (inverse of localization
length)  and gap energy as
a function of interchain hopping for $y=0.5$ and ${\cal D}=10^{-6}$.
For a given value of $\tilde{t}_\perp$ only the largest of the two
energies is represented.  
We note that for small interchain
hopping, there is a small \emph{reinforcement} of localization with
respect to the single chain limit. The origin of this reinforcement is
the development of $2k_F$ CDW$^\pi$ fluctuations in the system. With
stronger interchain hopping, delocalization effects become
apparent. Finally, when the interchain spin gap becomes larger than
the  $2k_F$ pinning energy, the system develops a $4k_F$ pinned CDW 
with a much smaller pinning energy. }

\label{fig:e_pin}
\end{figure}

In the case of two chains of  spinless
fermions\cite{orignac_spinless_ladder}, it is known that for repulsive
interactions there is a strong reinforcement of Anderson localization, 
but complete delocalization for attractive interactions. By contrast,
in the case of spinful fermions, 
the localization-delocalization transition is known to be obtained only for
$K_{\rho+}>3/2$ i. e. large
attractive interaction strength\cite{orignac_2chain_long}. Moreover, in
the case of repulsive interactions, in contrast with the spinless case,
there is no enhancement of localization for infinitesimal
disorder\cite{orignac_2chain_long} in the spinful system. This
behavior is due to the fact that both for repulsive and attractive
interactions, the localized phase in the presence of infinitesimal
disorder is a pinned $4k_F$ CDW.
Our RG study shows that when disorder becomes stronger (see figure
\ref{fig:pd}) or interchain hopping becomes weaker (see figure \ref{fig:pdtu}),
 a $2k_F$ pinned CDW$^\pi$ can be induced in the place of the $4k_F$ CDW. 
This pinned CDW$^\pi$ has a much shorter
localization length than the pinned $4k_F$ CDW. Interestingly, the
domain of existence of the PCDW$^\pi$ is larger for repulsive
interactions. Thus, for weak but not infinitesimal disorder,
localization will be reinforced by repulsive interactions. 
A similar effect is also obtained by reducing $t_\perp$. 
This results from the renormalization of
interactions in the $\sigma-$ sector by disorder, and could not be
obtained within  the infinitesimal disorder limit of
\cite{orignac_2chain_long}.   Such behavior
is represented on Fig. \ref{fig:e_pin} where the localization energy
$E_{\mathrm{loc.}}=\frac {v_F}{\xi_{\mathrm{loc.}}}$  and the gap energy
$E_{\mathrm{gap}}=\frac{v_F}{\xi_{\mathrm{gap}}}$
($\xi_{\mathrm{loc.}}=a e^{l_\mathrm{loc.}}$
and $\xi_{\mathrm{gap}}=a e^{l_\mathrm{gap}}$ being respectively the lengthscale at which
disorder or interaction become of order one) are represented a a function of
$\tilde{t}_\perp$ for fixed disorder and interaction. A slight
enhancement of the pinning energy can be observed for
$\tilde{t}_\perp$ large enough to deconfine but not strong enough to
make the gaps robust to disorder.

\begin{figure}[htbp]

\epsfig{file=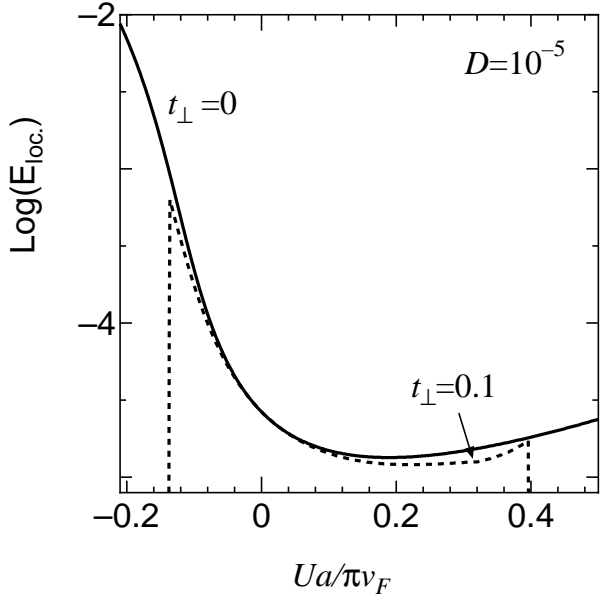,angle=0,width=8cm}
\caption{The behavior of the pinning energy (inverse localization %
length) a a function of interaction strength for $D=10^{-5}$ and
$t_\perp=0$ (single chain case, solid line) and $t_\perp=0.1$ (two %
chain case, dotted line). For weak interaction, the pinning energy is
the same for both systems. This corresponds to the confinement
regime. For stronger interaction, the two chain
system is always less localized than its single chain
counterpart. Thus, a large enough $t_\perp$ always induces delocalization. 
For an even stronger interaction, the two chain system goes into the
pinned $4k_F$ CDW phase with  a much longer pinning length. This limit
has been previously discussed in \cite{orignac_2chain_long}.} 
\label{fig:e_pin_U}
\end{figure}

Let us now turn to the question of the crossover from single chain to
two chain behavior. 
In the limit of $\tilde{t}_\perp=0$ i. e.  the single chain fermion system, 
it has been known for some time\cite{giamarchi_persistent_1d} that the
localization length was $\xi_{\mathrm{loc.}}\sim (1/D)^{1/(3-K_\rho)}$
for attractive interactions, but  \\ $\xi_{\mathrm{loc.}}\sim (1/D)^{1/(2-K_\rho)}$ for repulsive interactions, leading to a
reinforcement of localization by attractive interactions. This
reinforcement of localization results in a maximum of localization
length in the vicinity of the non-interacting point.  
For large $\tilde{t}_\perp$ and infinitesimal disorder, it has been shown in
\cite{orignac_2chain_long} that the localization length was
$\xi_{loc.}\sim (1/D)^{2/(3-2K_{\rho+})}$ so that the reinforcement of
localization by attractive interactions disappeared.  
In the present case of intermediate $\tilde{t}_\perp$ and non-infinitesimal
disorder, the behavior of the localization length deduced from the RG
is richer. For small interaction strength, disorder is strong enough
to prevent the formation of SCs or SCd phase in the system, and the
behavior of localization length is essentially one dimensional,
leading to a maximum of localization length in the vicinity of the
non-interacting point. For stronger interactions, the SCs and SCd
phases can develop, leading to a strong enhancement of localization
length caused by the weaker pinning of $4k_F$ CDWs compared to $2k_F$
CDWs. This result in the presence of two minimums of the localization
length as  a function of interaction. The corresponding behavior of
the pinning energy $E_{\mathrm{loc.}}=v_F/\xi_{\mathrm{loc.}}$ in
the case of ${\cal D}=10^{-5}$ and $Ua/(\pi v_F)=0.5$ is
represented on Fig. \ref{fig:e_pin_U} .Interestingly, the presence of
these minimums is asymmetric especially for small $\tilde{t}_\perp$. 
This behavior can be understood by looking in more details at the RG
equations. At a scale $l<l_{2ch.}$, where $l_{2ch.}$ is such that
$t_\perp(l_{2ch.})=\frac{\pi v_F}{a}$, the RG equations are those of
the one dimensional system, and interactions in the spin sector are
renormalized towards zero in the case of repulsive interactions. Only
for $l>l_{2ch.}$ are interactions in the spin sector growing again to
lead to the spin gap. By contrast, for attractive interactions, the
interactions are always growing in the spin sector, irrespective of
the strength of $\tilde{t}_\perp$. This implies that the spin gap is much
weaker in the case of small $\tilde{t}_\perp$ and repulsive interactions  thus
making the system  
more sensitive to disorder. Therefore, in the case of the ladder,
enhancement of localization is not determined only by the sign of
interaction but also by their strength. In the case of weak
interactions, the behavior is similar to the single chain case. For
stronger interactions, the effect can be inverted and attractive
interactions can have a delocalizing effect. Finally, for even
stronger interaction, the difference between attractive and repulsive
interactions becomes unobservable. It would be interesting to study
how the charge stiffness is modified by disorder in these different
regimes.    

A problem related to the single chain to two chain crossover is the
issue of confinement i.e. the irrelevance of single particle hopping
under RG. As discussed in section \ref{sec:confinement_transition},
the criterion for deconfinement is $E_{\mathrm{loc.}}=t_\perp$. This
criterion indeed reproduces the phase boundary between the phase III
(deconfined) and IV (confined) as can be seen by comparing
Figs. \ref{fig:pdtu} and \ref{fig:e_pin_U}. 
Interestingly, a kind of confinement  seems  to be
observed in phase (IIa) as a result of attractive interaction. However, 
 an effective 
two-particle hopping is known to be generated in this case
\cite{bourbonnais_houches,kishine_interchain} and plays the role of a
Josephson coupling between the chain. 
Such coupling  causes the formation of a gap in the transverse charge mode 
 so that the phase (IIa) is still the SCs phase in the absence of
disorder and becomes the $4k_F$ pinned CDW in the presence of disorder. 
Confinement of carriers by disorder  is obtained  in the case of strong
disorder. 
Such a confinement 
 has not been considered in the previous study of 
 the spinless fermions case\cite{orignac_spinless_ladder} 
 due to a choice of large $t_{\perp}$. 
 The  difference in confinement is as follows. 
 In the spinful case
and for a single chain, disorder generates a spin gap in the localized
phase. Thus, in the localized phase (region (IV) in Fig. 5), 
a gap in the single chain 
 has to be overcome to permit
single particle hopping. If one turns on a weak single particle hopping
between two chains, its effect is negligible  since the fields
$\phi_{\rho +}$ in both chains are  disordered as a result of the presence 
of strong disorder (this is to be contrasted with the phase IIa).   
In the case of spinless fermions, no spin degree of freedom is 
available to induce a spin gap so that the ``confinement'' results only
from the growth of the $y_\perp$ term under RG. 
 As one
can see from this discussion, this effect is not limited to the two
chain problem but is also present for any number of coupled chains. 
Therefore, confinement by disorder is a generic feature of coupled
disordered spinful chains and should also be observable for instance
with three coupled chains.

\section{Conclusion}
In the present paper, we have applied RG techniques to study Anderson
localization in a two-chain Hubbard ladder for arbitrary ratio of
$t_\perp,U,{\cal D}$. We have found that there
were three different types of localized phases: the $4k_F$ pinned CDW
(I and II) 
the $2k_F$ CDW$^\pi$ (III) and the decoupled single chain $2k_F$ CDW
(IV).  
We have been able to obtain the full phase diagram of this system. 
We have shown that a weak interchain hopping could lead to a
\emph{reinforcement} of localization, but a strong interchain hopping
resulted in a reduction of localization as discussed in
\cite{orignac_2chain_long}. The presence of $2k_F$ CDW phases should
have interesting consequences on transport properties that need
separate investigation. A remaining open problem is to study the
generation of the $4k_F$ disorder (i.e. disorder that couples to the
$4k_F$ CDW operator) directly from the RG equations and the resulting
overall behavior of conductivity. This will be left
for a future study.


\noindent

\begin{acknowledgement}
We thank S. Fujimoto, T. Giamarchi, 
N. Kawakami, M. Tsuchiizu and H. Yoshioka for
discussions. 
One of the authors (E. O.) is thankful for the financial support
from  Nagoya University where he stayed  from January to March in 2001. 
This work was partially supported by a
Grant-in-Aid for Scientific Research from the Ministry of
Education, Science, Sports and Culture (Grant No.09640429), Japan.
\end{acknowledgement}

\begin{table}
\caption{The two possible fixed points in a two-chain Hubbard ladder}

 \begin{tabular}{|c|c|c|c|}
 \hline
 $U>0$ & $\langle \theta_{\rho-}\rangle=0$ & $\langle
 \phi_{\sigma-}\rangle=\frac \pi 2$ & $\langle
 \phi_{\sigma+}\rangle=\frac \pi 2$  \\
 \hline 
 $U<0$ & $\langle \theta_{\rho-}\rangle=0$ & $ \langle
 \phi_{\sigma-}\rangle=0$ & $\langle
 \phi_{\sigma+}\rangle=0$\\  
 \hline
\end{tabular}

\label{tab:fixed-points}
\end{table}

\begin{table}
\caption{The fixed points of the disordered two-chain Hubbard ladder}

 \begin{tabular}{|c|c|c|c|c|}
 \hline
 I & $\langle \theta_{\rho-}\rangle=0$ & $\langle
  \phi_{\sigma-}\rangle=\frac \pi 2$ & $\langle
  \phi_{\sigma+}\rangle=\frac \pi 2$ & $t_\perp \to \infty$ \\
 \hline 
  IIa & $K_{\rho-} \simeq 1$ & $K_{\sigma-} \simeq 1$ & 
     $ \langle \phi_{\sigma+}\rangle=0$ & $t_\perp \to 0$ \\
 \hline
 IIb & $\langle \theta_{\rho-}\rangle=0$ & $ \langle
   \phi_{\sigma-}\rangle=0$ & $\langle
   \phi_{\sigma+}\rangle=0$& $t_\perp \to \infty$ \\  
 \hline
 III & $\langle \theta_{\rho-}\rangle=0$ & $ \langle \theta_{\sigma-}\rangle=0$     & $\langle \phi_{\sigma+}\rangle=0$& $t_\perp \to \infty$ \\
 \hline
 IV & $K_{\rho-} \simeq 1$ & $K_{\sigma-} \simeq 1$ & 
     $  \langle \phi_{\sigma+}\rangle=0$ & $t_\perp \to 0$ \\
 \hline
\end{tabular}

\label{tab:disordered-phases}
\end{table}

\end{document}